\documentclass[fleqn,10pt]{wlscirep}
\usepackage{amsmath}
\newtheorem{theorem}{Theorem}

\usepackage{bm}

\title{A Stronger Multi-observable Uncertainty Relation}

\author[1]{Qiu-Cheng Song}
\author[1,3]{Jun-Li Li}
\author[1]{Guang-Xiong Peng}
\author[1,2,3*]{Cong-Feng Qiao}
\affil[1]{Department of Physics, University of Chinese Academy of Sciences, YuQuan Road 19A, Beijing 100049, China}
\affil[2]{Department of Physics \& Astronomy, York University, Toronto, ON M3J 1P3, Canada}
\affil[3]{Key Laboratory of Vacuum Physics, University of Chinese Academy of Sciences}
\affil[*]{To whom correspondence should be addressed; E-mail: qiaocf@ucas.ac.cn}

\begin{abstract}
Uncertainty relation lies at the heart of quantum mechanics, characterizing the incompatibility of non-commuting observables in the preparation of quantum states. An important question is how to improve the lower bound of uncertainty relation. Here we present a variance-based sum uncertainty relation for $N$ incompatible observables stronger than the simple generalization of an existing uncertainty relation for two observables. Further comparisons of our uncertainty relation with other related ones for spin-$\frac{1}{2}$ and spin-$1$ particles indicate that the obtained uncertainty relation gives a better lower bound.
\end{abstract}
\begin{document}

\flushbottom
\maketitle
\thispagestyle{empty}

\section*{Introduction}
Uncertainty relation is one of the fundamental building blocks of quantum theory, and now plays an important role in quantum mechanics and quantum information \cite{PBusch,HHofmann,OGuhne,CAFuchs}. It is introduced by Heisenberg \cite{heis} in understanding how precisely the simultaneous values of conjugate observables could be in microspace, i.e., the position $X$ and momentum $P$ of an electron. Kennard \cite{Kennard} and Weyl \cite{Weyl} proved the uncertainty relation
\begin{eqnarray}\label{Robertson}
\Delta X\Delta P\geq
\frac{\hbar}{2}\; ,
\end{eqnarray}
where the standard deviation of an operator $X$ is defined by
$\Delta X =\sqrt{\langle\psi|X^2|\psi\rangle-\langle\psi|X|\psi\rangle^2}$.
Later, Robertson proposed the well-known formula of uncertainty relation \cite{Robertson}
\begin{eqnarray}\label{Robertson}
(\Delta A)^2(\Delta B)^2\geq
\left|\frac{1}{2}\langle\psi|[A,B]|\psi\rangle\right|^2\; ,
\end{eqnarray}
which is applicable to arbitrary incompatible observables, and the commutator is defined by $[A,B]=AB-BA$. The uncertainty relation was further strengthed by Schr\"odinger \cite{schrodinger} with the following form
\begin{eqnarray}\label{schrodinger}
(\Delta A)^2(\Delta B)^2\geq\left|\frac 12 \langle[A,B]\rangle\right|^2
+\left|\frac{1}{2}\langle\{A,B\}\rangle - \langle A\rangle\langle B\rangle\right|^2\; .
\end{eqnarray}
Here the commutator defined as $\{A,B\}\equiv AB+BA$.

It is realized that the traditional uncertainty relations may not fully capture the concept of incompatible observables as the lower bound could be trivially zero while the variances are not. An important question in uncertainty relation is how to improve the lower bound and immune from triviality problem \cite{mp,Coles}. Various attempts have been made to find stronger uncertainty relations. One typical kind of relation is that of Maccone and Pati, who derived two stronger uncertainty relations
\begin{eqnarray}
(\Delta A)^2 + (\Delta B)^2 & \geq &
\pm i\langle\psi|[A,B]|\psi\rangle+|\langle\psi|A \pm iB|\psi^\perp\rangle|^2, \label{Maccone} \\
(\Delta A)^2 + (\Delta B)^2 & \geq &
\frac 12|\langle\psi^\perp_{A+B}|A+B|\psi\rangle|^2=\frac 12[\Delta (A+B)]^2, \label{Maccone2p}
\end{eqnarray}
where $\langle \psi|\psi^\perp\rangle = 0$, $|\psi^\perp_{A+B}\rangle\propto(A+B-\langle A + B\rangle)|\psi\rangle$ , and the sign on the right-hand side of the inequality takes $+(-)$ while $ i\langle[A,B]\rangle$ is positive (negative). The basic idea behind these two relations is adding additional terms to improve the lower bound. Along this line, more terms \cite{bannur,sun,song}  and weighted form of different terms \cite{xiao,zhang} have been put into the uncertainty relations. It is worth mentioning that state-independent uncertainty relations can immune from triviality problem \cite{Huang,L1,L2,Branciard}. Recent experiments have also been performed to verify the various uncertainty relations \cite{xue,du,baek,feng}.

Besides the conjugate observables of position and momentum, multiple observables also exist, e.g., three component vectors of spin and angular momentum. Hence, it is important to find uncertainty relation for multiple incompatible observables. Recently, some three observables uncertainty relations were studied, such as Heisenberg uncertainty relation for three canonical observables \cite {Kechrimparis}, uncertainty relations for three angular momentum components \cite{dammeier}, uncertainty relation for three arbitrary observables \cite{song}.
Furthermore, some multiple observables uncertainty relations were proposed, which include multi-observable uncertainty relation in product \cite{qin,naihuan} and sum \cite{long,chenfei} form of variances. It is worth noting that Chen and Fei derived an variance-based uncertainty relation \cite{chenfei}
\begin{equation}\label{chen}
\sum_{i=1}^{N}(\Delta A_{i})^{2}\geq\frac{1}{N-2}\left\{\sum_{1\leq i<j\leq N}\left[\Delta (A_{i}+A_{j})\right]^2
-\frac{1}{(N-1)^2}\left[\sum_{1\leq i<j\leq N}\Delta (A_{i}+A_{j})\right]^2\right\}\; ,
\end{equation}
for arbitrary $N$ incompatible observables, which is stronger than the one such as derived from the uncertainty inequality for two observables \cite{mp}.

In this paper, we investigate variance-based uncertainty relation for multiple incompatible observables. We present a new variance-based sum uncertainty relation for multiple incompatible observables, which is stronger than an uncertainty relation from summing over all the inequalities for pairs of observables \cite{mp}. Furthermore, we compare the uncertainty relation with existing ones for a spin-$\frac{1}{2}$ and spin-$1$ particle, which shows our uncertainty relation can give a tighter bound than other ones.

\section*{Results}
\begin{theorem}
For arbitrary $N$ observables $A_{1}$, $A_{2}$, $\ldots$, $A_{N}$, the following variance-based uncertainty relation holds
\begin{equation}\label{relation}
\sum^N_{i=1}\Delta(A_i)^2\geq
\frac{1}{N}\left[\Delta(\sum^N_{i=1}A_i)\right]^2
+\frac{2}{N^2(N-1)}\left[\sum_{1\leq i<j\leq N}\Delta(A_i-A_j)\right]^2.
\end{equation}
The bound becomes nontrivial as long as the state is not common eigenstate of all the $N$ observables.
\end{theorem}

{\bf Proof:} To derive (\ref{relation}), start from the equality
\begin{equation}\label{relation1}
\sum_{1\leq i<j\leq N}\left[\Delta(A_{i}-A_{j})\right]^{2}=N \sum_{i=1}^{N}(\Delta A_{i})^{2}-\left[\Delta(\sum_{i=1}^{N}A_{i})\right]^{2},
\end{equation}
then using the inequality
\begin{equation}\label{relation2}
\frac{N(N-1)}{2}\sum_{1\leq i<j\leq N}\left[\Delta(A_{i}-A_{j})\right]^{2} \geq\left[\sum_{1\leq i<j\leq N}\Delta(A_{i}-A_{j})\right]^{2},
\end{equation}
we obtain the uncertainty relation (\ref{relation}).QED.

\begin{figure}[ht]
\centering
\includegraphics[width=0.4\textwidth]{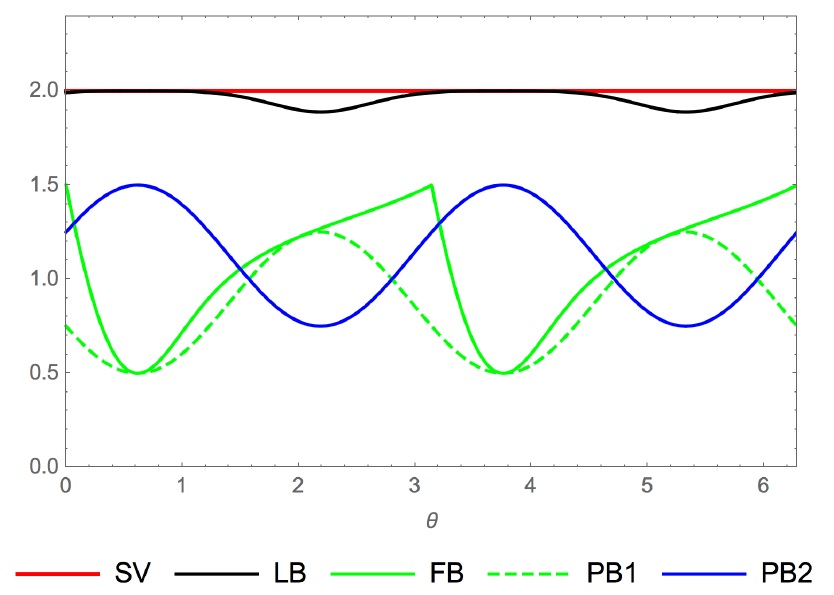}
\caption{(color online). Example of comparison between our relation (\ref{relation}) and the ones (\ref{chen}),(\ref{Maccone3p}),(\ref{Maccone3m}).  The upper line is the sum of the variances (SV) $(\Delta \sigma_x)^{2}+(\Delta \sigma_y)^{2}+(\Delta \sigma_z)^{2}$. The black line is the lower bound (LB) given by our relation (\ref{relation}).
The solid green line is the bound (\ref{chen}) (FB). The dashed green line is the bound (\ref{Maccone3p}) (PB1). The blue line is the bound (\ref{Maccone3m}) (PB2).}\label{f1}
\end{figure}

To show that our relation (\ref{relation}) has a stronger bound, we consider the result in Ref.~\citen{mp}, the relation (\ref{Maccone2p}) is derived from the uncertainty equality
\begin{equation}
\Delta A^{2}+\Delta B^{2}=\frac{1}{2}[\Delta(A+B)]^2+\frac{1}{2}[\Delta(A-B)]^2.
\end{equation}
Using the above uncertainty equality, one can obtain two inequalities for arbitrary $N$ observables, namely
\begin{equation}\label{Maccone3p}
\sum_{i=1}^{N}(\Delta A_{i})^{2}\geq\frac{1}{2(N-1)}\sum_{1\leq i<j\leq N}[\Delta(A_{i}+A_{j})]^{2}
\end{equation}
and
\begin{equation}\label{Maccone3m}
\sum_{i=1}^{N}(\Delta A_{i})^{2}\geq\frac{1}{2(N-1)}\sum_{1\leq i<j\leq N}[\Delta(A_{i}-A_{j})]^{2}.
\end{equation}
The bound in (\ref{chen}) is tighter than the one in (\ref{Maccone3p}) \cite{chenfei}. However, the lower bound in (\ref{chen}) is  not always tighter than the one in (\ref{Maccone3m}) (see Figure \ref{f1}).

{\bf Example 1} To give an overview that the relation (\ref{relation}) has a better lower bound than the relations (\ref{chen}),(\ref{Maccone3p}),(\ref{Maccone3m}), we consider a family of qubit pure states given by the Bloch vector
$\vec{r}=(\frac{1}{\sqrt{2}}\cos\theta,\frac{1}{\sqrt{2}}\cos\theta,\sin\theta)$, and choose the Pauli matrices
\begin{equation*}
\sigma_{x}=\begin{pmatrix}
0 & 1\\1 & 0
\end{pmatrix},~~~
\sigma_{y}=\begin{pmatrix}
0 & -i\\i & 0
\end{pmatrix},~~~
\sigma_{z}=\begin{pmatrix}
1 & 0\\0 & -1
\end{pmatrix}.
\end{equation*}
Then we have $(\Delta \sigma_{x})^{2}+(\Delta \sigma_{y})^{2}+(\Delta \sigma_{z})^{2}=2$,
$[\Delta( \sigma_{x}+ \sigma_{y})]^{2}=2(\sin\theta)^2$,
and $[\Delta( \sigma_{y}+ \sigma_{z})]^{2}=[\Delta( \sigma_{x}+ \sigma_{z})]^{2}
=\frac{5}{4}+\frac{1}{4}\cos2\theta-\frac{\sqrt{2}}{2}\sin2\theta$.
Similarly, $[\Delta( \sigma_{x}- \sigma_{y})]^{2}=2$, and $[\Delta( \sigma_{y}- \sigma_{z})]^{2}=[\Delta( \sigma_{x}- \sigma_{z})]^{2}=\frac{5}{4}+\frac{1}{4}\cos2\theta+\frac{\sqrt{2}}{2}\sin2\theta$.
The comparison between the lower bounds (\ref{chen}),(\ref{Maccone3p}),(\ref{Maccone3m}) and (\ref{relation})
is given in Figure \ref{f1}.
Apparently, our bound is tighter than (\ref{chen}) ,(\ref{Maccone3p}) and (\ref{Maccone3m}). We shall show with detailed proofs and examples that our uncertainty relation (\ref{relation}) has better lower bound than that of (\ref{chen}),(\ref{Maccone3p}),(\ref{Maccone3m}) in the following sections.

\subsection*{Comparison between the lower bound of our uncertainty relation (\ref{relation}) with that of inequality (\ref{Maccone3p})}

First, we compare our relation (\ref{relation}) with the one (\ref{Maccone3p}).
Note that $\Delta(A_i+A_j)^2=\Delta A_i^2+\Delta A_j^2+\langle \{A_i,A_j\}\rangle
-2\langle A_i\rangle\langle A_j\rangle$,
the relation (\ref{Maccone3p}) becomes
\begin{equation}
\sum_{i=1}^{N}(\Delta A_{i})^{2}
\geq\frac{1}{2(N-1)}\left\{(N-1)\sum_{i=1}^{N}(\Delta A_{i})^{2}
+\sum_{1\leq i<j\leq N}[\langle\{A_i,A_j\}\rangle-2\langle A_i\rangle\langle A_j\rangle]\right\}.
\end{equation}
Simplify the above inequality, we obtain
\begin{equation}\label{patis}
\sum_{i=1}^{N}(\Delta A_{i})^{2}
\geq\frac{1}{(N-1)}\sum_{1\leq i<j\leq N}[\langle\{A_i,A_j\}\rangle-2\langle A_i\rangle\langle A_j\rangle],
\end{equation}
which is equal to the relation (\ref{Maccone3p}).

Similarly, by using $\Delta(A_i-A_j)^2=\Delta A_i^2+\Delta A_j^2-\langle \{A_i,A_j\}\rangle
+2\langle A_i\rangle\langle A_j\rangle$,  our relation (\ref{relation}) becomes
\begin{equation}
\begin{split}
&\sum_{i=1}^{N}(\Delta A_{i})^{2}\geq
\frac{1}{N}\left[\sum_{i=1}^{N}(\Delta A_{i})^{2}
+\sum_{1\leq i<j\leq N}[\langle\{A_i,A_j\}\rangle-2\langle A_i\rangle\langle A_j\rangle]\right]\\
&+\frac{2}{N^2(N-1)}\left\{(N-1)\sum_{i=1}^{N}(\Delta A_{i})^{2}
-\sum_{1\leq i<j\leq N}[\langle\{A_i,A_j\}\rangle-2\langle A_i\rangle\langle A_j\rangle]
+\sum^{i\neq i' or j\neq j'}_{\substack{1\leq i<j\leq N\\1\leq i'<j'\leq N}}\Delta(A_{i}-A_{j})\Delta(A_{i'}-A_{j'})\right\}.
\end{split}
\end{equation}
Simplify the above inequality, we get
\begin{equation}\label{relations}
\sum_{i=1}^{N}(\Delta A_{i})^{2}\geq
\frac{1}{N-1}\sum_{1\leq i<j\leq N}[\langle\{A_i,A_j\}\rangle-2\langle A_i\rangle\langle A_j\rangle]
+\frac{2}{(N-2)(N^2-1)}\sum^{i\neq i' or j\neq j'}_{\substack{1\leq i<j\leq N\\1\leq i'<j'\leq N}}\Delta(A_{i}-A_{j})\Delta(A_{i'}-A_{j'}),
\end{equation}
which is equal to the relation (\ref{relation}). It is easy to see that the right-hand side of (\ref{relations}) is greater than the right-hand side of (\ref{patis}). Hence, the relation (\ref{relation}) is stronger than the relation (\ref{Maccone3p}).

\subsection*{Comparison between the lower bound of our uncertainty relation (\ref{relation}) with that of inequalities (\ref{chen}) and (\ref{Maccone3m})}

Here, we will show the uncertainty relation (\ref{relation}) is stronger than inequalities (\ref{Maccone3m}) and (\ref{chen}) for a spin-$\frac{1}{2}$ particle and measurement of Pauli-spin operators $\sigma_x,\sigma_y,\sigma_z$.
 Then the uncertainty relation (\ref{relation}) has the form
\begin{equation}\label{relation3}
(\Delta \sigma_{x})^{2}+(\Delta \sigma_{y})^{2}+(\Delta \sigma_{z})^{2}
\geq\frac{1}{3}\left[\Delta(\sigma_{x}+\sigma_{y}+\sigma_{z})\right]^{2}
+\frac{1}{9}\left(\sum_{1\leq i<j\leq 3}\Delta(\sigma_{i}-\sigma_{j})\right)^{2},
\end{equation}
the relation (\ref{Maccone3m}) is given by
\begin{equation}\label{Maccone3}
(\Delta \sigma_{x})^{2}+(\Delta \sigma_{y})^{2}+(\Delta \sigma_{z})^{2}
\geq\frac{1}{4}\sum_{1\leq i<j\leq 3}[\Delta(\sigma_{i}-\sigma_{j})]^{2},
\end{equation}
and the relation (\ref{chen}) says that
\begin{equation}\label{chen3}
(\Delta \sigma_{x})^{2}+(\Delta \sigma_{y})^{2}+(\Delta \sigma_{z})^{2}
\geq\sum_{1\leq i<j\leq 3}\left[\Delta (\sigma_{i}+\sigma_{j})\right]^2-\frac{1}{4}\left(\sum_{1\leq i<j\leq 3}\Delta (\sigma_{i}+\sigma_{j})\right)^2.
\end{equation}
We consider a qubit state
and its Bloch sphere representation
\begin{align}
\rho=\frac{1}{2}(I+\vec{r}\cdot\vec{\sigma}),
\end{align}
where $\vec{\sigma}=(\sigma_x,\sigma_y,\sigma_z)$ are Pauli matrices
and the Bloch vector $\vec{r}=(x,y,z)$ is real three-dimensional vector
such that $\|\vec{r}\|\leq 1$.
Then we have $(\Delta \sigma_{x})^{2}=\text{Tr}[\rho \sigma_{x} \sigma_{x} ]-\text{Tr}[\rho \sigma_{x}]^2=1-x^2$,
$(\Delta \sigma_{x})^{2}+(\Delta \sigma_{y})^{2}+(\Delta \sigma_{z})^{2}=3-(x^2+y^2+z^2)$.
The relation (\ref{relation3}) has the form
\begin{equation}\label{relation33}
(\Delta \sigma_{x})^{2}+(\Delta \sigma_{y})^{2}+(\Delta_\rho \sigma_{z})^{2}
\geq\frac{1}{9}\alpha^2+\frac{1}{3} \left(3-(x+y+z)^2\right),
\end{equation}
where we define $\alpha=\sqrt{2-(x-y)^2}+\sqrt{2-(x-z)^2}+\sqrt{2-(y-z)^2}$. And the relation (\ref{Maccone3}) becomes
\begin{equation}\label{Maccone33}
(\Delta \sigma_{x})^{2}+(\Delta \sigma_{y})^{2}+(\Delta \sigma_{z})^{2}
\geq\frac{1}{2} \left(3-(x^2+y^2+z^2)+xy+xz+yz\right).
\end{equation}
Let us compare the lower bound of (\ref{relation33}) with that of (\ref{Maccone33}). The difference of these two bounds is
\begin{align}\label{spati1}
&\frac{1}{9}\alpha^2
+\frac{1}{6}\left(x^2+y^2+z^2\right)-\frac{7}{6}\left(xy+xz+zy\right)-\frac{1}{2}\\
\geq&\left(\sqrt{2-(x-y)^2}\sqrt{2-(x-z)^2}\sqrt{2-(y-z)^2}\right)^{\frac{2}{3}}-\left(x^2+y^2+z^2\right)-\frac{1}{2}\notag\\
\geq&\left(\sqrt{2-(x-y)^2}\sqrt{2-(x-z)^2}\sqrt{2-(y-z)^2}\right)^{\frac{2}{3}}-\frac{3}{2}\notag,
\end{align}
for all $x,y,z\in[-1,1]$. When $x=y=z=\pm1/\sqrt{3}$, the above inequality becomes equality,
then the Eq.(\ref{spati1}) has the minimum value $1/2>0$.
This illustrates that the uncertainty relation (\ref{relation}) is stronger that the one (\ref{Maccone3m}) for a spin-$\frac{1}{2}$ particle and measurement of Pauli-spin operators $\sigma_x,\sigma_y,\sigma_z$.

Let us compare the uncertainty relation (\ref{relation3}) with (\ref{chen3}). The relation (\ref{chen3}) has the form
\begin{equation}\label{chen33}
(\Delta \sigma_{x})^{2}+(\Delta \sigma_{y})^{2}+(\Delta \sigma_{z})^{2}
\geq 6-2(xy+xz+yz)-2(x^2+y^2+z^2)-\frac{1}{4}\beta^2,
\end{equation}
where we define $\beta=\sqrt{2-(x+y)^2}+\sqrt{2-(x+z)^2}+\sqrt{2-(y+z)^2}$.
Then the difference of these two bounds of relation (\ref{relation33}) and (\ref{chen33}) becomes
\begin{align}\label{sfei2}
&\frac{1}{9}\alpha^2+\frac{1}{4}\beta^2+\frac{2}{3}(x+y+z)^2+(x^2+y^2+z^2)-5\\
\geq&\left(\frac{\sqrt{6}}{9}\alpha+\frac{\sqrt{2}+\sqrt{6}}{4}\beta\right)^2
\left(\frac{3}{8+3\sqrt{3}}\right)
+\frac{1}{2}(x-z)^2-5\notag,
\end{align}
where we have twice used Cauchy's inequality. When $\alpha=\frac{\sqrt{6}}{3},\beta=\frac{\sqrt{2}+\sqrt{6}}{2},x+y+z=0$ and
$x=-y=\pm1/\sqrt{2},z=0~ or ~x=-z=\pm1/\sqrt{2},y=0 ~or ~y=-z=\pm1/\sqrt{2},x=0$, the above inequality becomes equality,
then the Eq.(\ref{sfei2}) has the minimum value $\sqrt{3}-\frac{4}{3}>0$.
This illustrates that the uncertainty relation (\ref{relation}) is stronger that the one (\ref{chen})
for a spin-$\frac{1}{2}$ particle and measurement of Pauli-spin operators $\sigma_x,\sigma_y,\sigma_z$.

\begin{figure}[ht]
\centering
\includegraphics[width=0.8\textwidth]{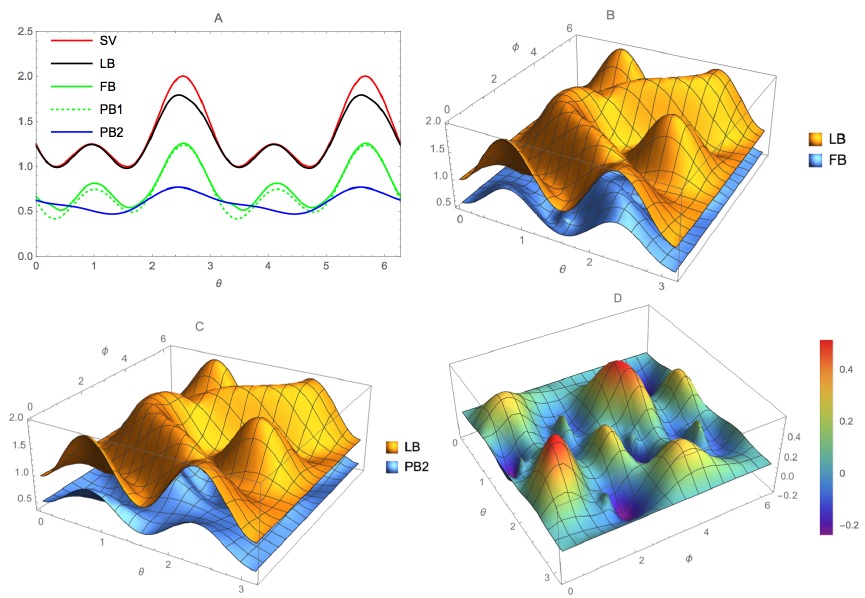}
\caption{(color online). Example of comparison between our relation (\ref{relation}) and ones (\ref{chen}),(\ref{Maccone3p}),(\ref{Maccone3m}).
We choose $A_1=L_x$, $A_2=L_y$ and $A_3=L_y$ three components of the angular momentum for spin-$1$ particle,
and a family of states parametrized by $\theta$ and $\phi$ as
$|\psi\rangle=\sin \theta \cos \phi |1\rangle+\sin \theta  \sin \phi |0\rangle+\cos \theta |-1\rangle$.
[A] For $\phi=\pi/4$, the comparison between our relation (\ref{relation}) and ones (\ref{chen}),(\ref{Maccone3p}),(\ref{Maccone3m}). The upper line is the sum of the variances (SV) $(\Delta L_x)^{2}+(\Delta L_y)^{2}+(\Delta L_z)^{2}$. The black line is the lower bound (LB) given by our relation (\ref{relation}).
The solid green line is the bound (\ref{chen}) (FB). The dashed green line is the bound (\ref{Maccone3p}) (PB1). The blue line is the bound (\ref{Maccone3m}) (PB2).
[B] The comparison between our relation (\ref{relation}) and (\ref{chen}), which shows that our relation (\ref{relation}) (LB) has stronger bound than (\ref{chen}) (FB).
[C] The comparison between our relation (\ref{relation}) and (\ref{Maccone3m}), which shows that our relation (\ref{relation}) (LB) has stronger bound than (\ref{Maccone3m}) (PB2).
[D] The lower bound of the relation (\ref{chen}) minus the lower bound of the relation (\ref{Maccone3m}). }\label{f2}
\end{figure}
\vspace{.1cm}

{\bf Example 2} For spin-1 systems, we consider the following quantum state characterized by $\theta$ and $\phi$
\begin{align}
|\psi\rangle=\sin \theta \cos \phi |1\rangle
+\sin \theta  \sin \phi |0\rangle+\cos \theta |-1\rangle,
\end{align}
with $~0\leq\theta\leq\pi,~0\leq\phi\leq2\pi$. By choosing the three angular momentum operators ($\hbar=1$)
\begin{equation*}
L_{x}=\frac{1}{\sqrt{2}}\begin{pmatrix}
0 & 1 & 0\\1 & 0 & 1\\0 & 1 & 0
\end{pmatrix},~~~
L_{y}=\frac{1}{\sqrt{2}}\begin{pmatrix}
0 & -i & 0\\i & 0 & -i\\0 & i & 0
\end{pmatrix},~~~
L_{z}=\begin{pmatrix}
1 & 0 & 0\\0 & 0 & 0\\0 & 0 & -1
\end{pmatrix},
\end{equation*}
the comparison between the lower bounds (\ref{chen}),(\ref{Maccone3p}),(\ref{Maccone3m}) and (\ref{relation})
is shown by Figure \ref{f2}. The results suggest that the relation (\ref{relation})
can give tighter bounds than other ones  (\ref{chen}),(\ref{Maccone3p}),(\ref{Maccone3m})
for a spin-$1$ particle and measurement of angular momentum operators $L_x$, $L_y$, $L_z$.

\section*{Conclusion}

We have provided a variance-based sum uncertainty relation for $N$ incompatible observables, which is stronger than the simple generalizations of the uncertainty relation for two observables derived by Maccone and Pati [Phys. Rev. Lett. {\bf113}, 260401 (2014)]. Furthermore, our uncertainty relation gives a tighter bound than the others by comparison for a spin-$\frac{1}{2}$ particle with the measurements of spin observables $\sigma_x,\sigma_y,\sigma_z$. And also, in the case of spin-$1$ with measurement of angular momentum operators $L_x,L_y,L_z$,
our uncertainty relation predicts a tighter bound than other ones.


\section*{Acknowledgements}
This work was supported in part by the Ministry of Science and Technology of the People's Republic of China(2015CB856703); by the Strategic Priority Research Program of the Chinese Academy of Sciences, Grant No.XDB23030100; and by the National Natural Science Foundation of China(NSFC) under the grants 11375200 and 11635009.

\section*{Author contributions statement}
Q.-C. S. and J.-L. L. and G.-X. P. and C.-F. Q. contribute equally to this work, and agree with the manuscript submitted.

\section*{Additional information}
Competing financial interests: The authors declare no competing financial interests.

\end{document}